# Hidden structural transition in epitaxial $Ca_{0.5}Sr_{0.5}IrO_3$/$SrTiO_3$ thin film

W. T. Jin,[1] H. Gretarsson,[1] S. Y. Jang,[2,3] Chang-Yong Kim,[4] T. W. Noh,[2, 3] and Young-June Kim[1,*]

[1]Department of Physics, University of Toronto, 60 St. George Street, Toronto, M5S 1A7 Ontario, Canada
[2]Center for Correlated Electron Systems (CCES), Institute for Basic Science (IBS), Seoul 08826, Republic of Korea
[3]Department of Physics and Astronomy, Seoul National University, Seoul 08826, Republic of Korea
[4]Canadian Light Source, Saskatoon, S7N 0X4, Saskatchewan, Canada

E-mail: yjkim@physics.utoronto.ca

**Abstract**

A structural transition in an $ABO_3$ perovskite thin film involving the change of the $BO_6$ octahedral rotation pattern can be hidden under the global lattice symmetry imposed by the substrate and often easily overlooked. We carried out high-resolution x-ray diffraction experiments to investigate the structures of epitaxial $Ca_{0.5}Sr_{0.5}IrO_3$ (CSIO) perovskite iridate films grown on the $SrTiO_3$ (STO) and $GdScO_3$ (GSO) substrates in detail. Although the CSIO/STO film layer displays a global tetragonal lattice symmetry evidenced by the reciprocal space mapping, synchrotron x-ray data indicates that its room temperature structure is monoclinic due to Glazer's $a^+a^-c^-$-type rotation of the $IrO_6$ octahedra. In order to accommodate the lower-symmetry structure under the global tetragonal symmetry, the film breaks into 4 twinned domains, resulting in the splitting of the (half-integer, 0, integer) superlattice reflections. Surprisingly, the splitting of these superlattice reflections decrease with increasing temperature, eventually disappearing at $T_S$ = 510(5) K, which signals a structural transition to an orthorhombic phase with $a^+a^-c^0$ octahedral rotation. In contrast, the CSIO/GSO film displays a stable monoclinic symmetry with $a^+b^-c^-$ octahedral rotation, showing no structural instability caused by the substrate up to 520 K. Our study illustrates the importance of the symmetry in addition to the lattice mismatch of the substrate in determining the structure of epitaxial thin films.

## 1. Introduction

In the past decades, strain engineering has emerged as a powerful way to tune the structural and physical properties of the $ABO_3$ perovskite thin films and create novel functionalities in them.[1-3] Different from the hydrostatic pressure or uniaxial pressure generally applied for bulk systems, the biaxial epitaxial strain from the lattice-mismatched substrate offers a unique type of external perturbation to the thin film. The lattice mismatch between the substrate and the film-layer material is compensated by structural changes involving rotations or distortions of the $BO_6$ octahedra in the film layer. By changing the free energy landscape of the film-layer material, the epitaxial strain can lead to modification of the stable structures at different temperatures and tune the transitions between different structural phases accordingly. For instance, compressively strained $SrRuO_3$ thin film grown on the $SrTiO_3$ substrate displays an orthorhombic-to-tetragonal transition at the temperature ~ 270 K lower than that for bulk $SrRuO_3$, while such a transition was not observed for the tensilely strained $SrRuO_3$ films grown on the $DyScO_3$ and $GdScO_3$ substrates.[4, 5]

However, the global lattice symmetry of the substrate sometimes prevents epitaxial thin films from taking on the specific octahedral rotation pattern that minimizes free energy.[6,7] In this case, the film-layer material can exhibit local symmetry that is very different from the global symmetry. Together these domains can preserve the global lattice symmetry, while keeping lower local symmetry of individual domains. Such a decoupling between the internal octahedral rotation symmetry and average lattice symmetry was reported for the $SrTiO_3/KTaO_3$ thin film and $SrRuO_3/SrTiO_3$ thin film.[8,9] Due to the competition between the constraint from the substrate and the intrinsic tendency of the film-layer material towards new structural phases, it might end up with the formation of multiple domains in the film layer. Moreover, this type of distinction between local and global symmetry could hide the structural transition in the film-layer material resulting from the change of octahedral rotation pattern. Unveiling such hidden structural properties and figuring out the true internal symmetry will be crucial for understanding the physical properties of the thin film.

Here we present our comprehensive study of the structures of epitaxial $Ca_{0.5}Sr_{0.5}IrO_3$ films grown on $SrTiO_3$ (001) (in cubic notation, space group: *P m -3 m*[10]) and $GdScO_3$ (110) substrates (in orthorhombic notation, space group *P b n m*[11]) using x-ray diffraction. Thin films of perovskite iridates with strong spin-orbit coupling have attracted much attention recently, due to the high degrees of tunability of their physical properties and various possible topological phases.[12-20] As they lie in proximity to the metal-insulator phase boundary, their electronic properties are highly sensitive to subtle structural modulations caused by external perturbations, such as chemical pressure or epitaxial strain.[21-25] Previous experimental studies indicated that the $Ca_{0.5}Sr_{0.5}IrO_3$ film grown on GSO (110) substrate shows a nearly metallic behavior, whereas the $Ca_{0.5}Sr_{0.5}IrO_3$ film grown on STO (001) substrate displays an insulating behavior.[26] It was suggested that the different electronic properties of $Ca_{0.5}Sr_{0.5}IrO_3$ films grown on the GSO and STO substrates originate from the subtle differences in their crystal and band structures caused by the tensile and compressive epitaxial strain, respectively. Although $Ca_{0.5}Sr_{0.5}IrO_3$ does not exist in the bulk form due to the difficulty in obtaining the pure perovskite phase at the doping level of 50%, the unstrained lattice constants can be estimated to be *a* = 5.503 Å, *b* = 5.575 Å, *c* = 7.809 Å in the orthorhombic/monoclinic notation or $a_{pc}$ = 3.913 Å in pseudocubic notation, by performing the linear extrapolations of the derived lattice constants of perovskite $Ca_{1-x}Sr_xIrO_3$ (0.6≤ x≤1).[27] Compared with $a_{pc}$ = 3.905 Å for STO[10] and $a_{pc}$ = 3.970 Å for GSO[11], we can estimated the epitaxial strain to be -0.2% (compressive) for CSIO/STO and +1.5% (tensile) for CSIO/GSO, respectively. Through our high-resolution x-ray diffraction studies, the CSIO/STO and CSIO/GSO films were found to show distinct structural properties. Although the CSIO/STO film displays global tetragonal lattice symmetry, hinted by the change of diffraction pattern of superstructure reflections, a hidden monoclinic-to-orthorhombic structural transition at 510(5) K was revealed in the CSIO/STO film, which is however absent in the CSIO/GSO film up to 520 K. The symmetry of the substrate is found to play a key role in determining the structure of the film layer.

## 2. Experimental Section

High-quality CSIO thin films were epitaxially deposited on STO (001) and GSO (110) substrates, respectively, by pulsed laser deposition (PLD) technique as described in Ref. 23. Reciprocal space mappings of both samples were collected at the room temperature using the laboratory-based four-circle x-ray diffractometer, which employs a copper x-ray tube, a graphite monochromator, and a sodium iodide scintillation detector. Synchrotron x-ray diffraction experiments were performed at the Hard X-ray MicroAnalysis (HXMA) beamline at the Canadian Light Source (CLS) to study the structural properties of both films. The energy of incident x-ray was tuned to be 11.015 keV, far below the Ir $L_3$ absorption edge (11.215 keV). The x-ray beam was monochromatized by a Si(111) crystal and focused by a Pt-coated toroidal mirror before being scattered by the samples. The samples were mounted using the silver epoxy onto a homemade furnace stage with x-ray transparent shielding, which is capable of reaching a high temperature above 500 K. Either a pin diode or a cyberstar scintillation detector was used depending on the intensity of the diffracted beam.

## 3. Results and discussions

The interference fringes as a result of good film quality were clearly observed for both the CSIO/STO and CSIO/GSO films, from which the thicknesses of both films were estimated to be around 40 nm.[26] For convenience, the pseudocubic notation will be used for the CSIO films grown on both substrates throughout this paper, and the orientations of $a_{pc}$, $b_{pc}$ and $c_{pc}$ in the pseudocubic unit cell are shown in Fig. 1.

The reciprocal space maps of the films obtained at room temperature around the (1 0 3), (0 1 3), (-1 0 3) and (0 -1 3) symmetrical reflections of the STO substrate (cubic, space group: *P m -3 m*[10]), as well as those around the (3 3 2), (4 2 0), (3 3 -2) and (2 4 0) asymmetrical reflections of the GSO substrate (orthorhombic, space group: *P b n m*[11]) are shown in Fig. 2(a) and (b). The diffraction spots of both films possess the same $Q_x$ (in-plane wave vector) as the substrates, confirming the epitaxial nature of the CSIO films on both substrates. For CSIO/STO, the tetragonal symmetry of the CSIO layer is observed, similar to the in-plane tetragonality owned by STO substrate, as indicated by the same $Q_z$ (out-of-plane wave vector) for the {1 0 3} reflections. In contrast, the CSIO layer grown on the GSO substrate exhibits an orthorhombic/monoclinic unit cell, as suggested by the different $Q_z$ values of the {1 0 3} reflections.

As the lattice mismatch between the film layer and the substrate is generally accommodated by the distortions or rotations of the oxygen octahedra in the film layer, half-integer superstructure reflections in pseudocubic notation are expected.[6] The extinction rules of half-integer reflections corresponding to different types of octahedral rotations are summarized in Table 1. Glazer pointed out that there exist 23 possible rotation patterns of the $BO_6$ octahedra about three crystallographic axes of the pseudocubic cell, giving rise to different symmetries of the perovskite $ABO_3$ compounds.[7] For instance, no half-integer reflections are expected for $SrTiO_3$ in the cubic phase with all the $TiO_6$ octahedra unrotated (with $a^0a^0a^0$ rotation pattern in the Glazer notation), while the (odd, odd, odd)/2 superstructure reflections with H ≠ K will show up in the tetragonal phase of STO below 105 K arising from the $a^0a^0c^-$ type octahedral rotation, with adjacent $TiO_6$ octahedra rotating out-of-phase about the $c_{pc}$ axis.[28]

Table 1: The extinction rules of half-integer reflections in pseudocubic notation corresponding to different types of octahedral rotations.[7] "+" and "-" signify the in-phase and out-of-phase rotation of octahedra about specific crystallographic axis, respectively.

| Type of rotation | Allowed half-integer reflections | Restrictions |
|---|---|---|
| $a^+$ | (even odd odd)/2 | K ≠ L |
| $a^-$ | (odd odd odd)/2 | K ≠ L |
| $b^+$ | (odd even odd)/2 | H ≠ L |
| $b^-$ | (odd odd odd)/2 | H ≠ L |
| $c^+$ | (odd odd even)/2 | H ≠ K |
| $c^-$ | (odd odd odd)/2 | H ≠ K |

Fig. 2(c) and (d) show the L-scans along the (0, 1/2, L), (1/2, 0, L), (1/2, 1/2, L) and (1/2, 3/2, L) directions at room temperature. For both films, the appearance of the (0, 1/2, half-integer) superstructure peaks implies the in-phase rotations of the $IrO_6$ octahedra about the $a_{pc}$ axis, which can be denoted as $a^+$ rotation in the Glazer notation. The absence of (1/2, 3/2, integer) peaks rules out the possibility of $c^+$ rotation (the $IrO_6$ octahedra rotating in-phase about the $c_{pc}$ axis), and the presence of (1/2, 1/2, half-integer) peaks suggests the out-of-phase rotations of the $IrO_6$ octahedra about the $b_{pc}$ axis ($b^-$ rotation). In addition, superstructure peaks at (1/2, 0, integer) were also observed, which are commonly ascribed to the combination effect of $a^+$ and $b^-$ rotations acting in concert.[29, 30] Based on these superstructure reflections, the rotation patterns of the $IrO_6$ octahedra can be deduced to be $a^+a^-c^-$ ($a_{pc} = b_{pc}$, with equivalent magnitudes of rotations about $a_{pc}$ and $b_{pc}$) and $a^+b^-c^-$ ($a_{pc} \neq b_{pc}$, with inequivalent magnitudes of rotations about $a_{pc}$ and $b_{pc}$), for CSIO/STO and CSIO/GSO, respectively,[31] giving rise to the space group of $P112_1/m$ (monoclinic) for both films at room temperature.[7] For CSIO/STO, the equivalence of $a_{pc}$ and $b_{pc}$ ($a_{pc} = b_{pc}$) and equivalent magnitudes of rotations about these two axes are forced upon by the in-plane tetragonal lattice of the STO substrate. In contrast, for CSIO/GSO, the orthorhombic substrate requires the inequivalence of $a_{pc}$ and $b_{pc}$ ($a_{pc} \neq b_{pc}$) and inequivalent magnitudes of rotations about them are allowed naturally. In the discussions below, further evidence supporting such a monoclinic space group for the room-temperature structure of both films will be presented. The monoclinic symmetry due to octahedral rotations in both films is consistent with the a recent structural study on SrIrO$_3$ (SIO) films grown on the STO and GSO substrates, which revealed that either the compressive or tensile strain will distort the angle γ of the SIO film layer to deviate from 90°.[32]

For the CSIO/GSO film, the monoclinic octahedral rotation symmetry is consistent with the orthorhombic/monoclinic lattice symmetry suggested by the RSM as shown in Fig. 2(b). A single-domain monoclinic structure with $a^+b^-c^-$ -type octahedral rotation labeled A' is suggested by Fig. 2(d). However, for the CSIO/STO film, such a monoclinic symmetry seem to contradict the tetragonal lattice of the film layer suggested by the RSM as shown in Fig. 2(a). The only way to accommodate such a monoclinic or an orthorhombic distortion in a tetragonal lattice is to incorporate multiple low-symmetry domains. Indeed, for CSIO/STO, the (1/2, 0, half-interger) and (0, 1/2, integer) peaks were also observed (marked as B in Fig. 2(c)), indicating the existence of a second domain with the $a^-a^+c^-$- type tilt, in which the $a_{pc}$ and $b_{pc}$ axes are exchanged with respect to the first domain. Therefore, the octahedral rotation pattern is $a^+a^-c^-$ and $a^-a^+c^-$ for the two domains existing in the CSIO/STO film layer, labeled A and B respectively.

For the CSIO/STO film, K-scans at the room temperature across the superstructure peaks from domain A indicates that there are in fact two twinned subdomains within this domain. As shown in Fig. 3(a), the superstructure reflection (1/2, 0, 3) split into two well-defined peaks along the K direction. We can rule out the possibility of these peaks arising from incommensurability because the splitting between the two peaks (ΔK) diminishes with decreasing L and is proportional to L (ΔK ∝ L), as shown in the inset of Fig. 3(a). The observed L dependence of ΔK suggests that the splitting occurs in angle, not in Q, which can arise from twinning within domain A. On the other hand, a similar case was observed for the superstructure peaks from domain B (Fig. 3(b)), showing an L-dependent peak splitting along the H direction. Note that the peak splitting is indiscernible for L = 1, as plotted in the inset of Fig. 3(b), due to the limited Q-resolution of the detector.

To account for the observed splitting of superstructure reflections for the CSIO/STO film, a 4-domain structure with monoclinic distortions at room temperature is proposed and illustrated in Fig. 4(a). As shown in Fig. 2(c), domain A and domain B, with $a^+a^-$- and $a^-a^+$-type octahedral rotation, respectively, coexist in the CSIO/STO film layer. The splitting of (1/2, 0, 3) peak along the K direction (see Fig. 3) suggests that the angle between the pseudocubic $b_{pc}$ and $c_{pc}$ axes deviates from 90° (α ≠ 90°) in domain A, leading to two monoclinic subdomains (A1 and A2) twinned with respect to the *ac* plane. Due to the constraint from the STO substrate which requires the in-plane tetragonal lattice ($a_{pc} = b_{pc}$), as indicated by Fig. 2(a), the only possible rotation pattern of the CSIO film layer is $a^+a^-c^-$, with the IrO$_6$ octahedron rotating out-of-phase around $c_{pc}$. Similarly, the splitting of (0, 1/2, 3) peak along the H direction suggests two monoclinic subdomains (B1 and B2, β ≠ 90°) twinned with respect to the *bc* plane within domain B, with the octahedral rotation pattern of $a^-a^+c^-$. Therefore, a total of 4 monoclinic subdomains with the space group of *P*112$_1$/*m* can explain the observed splitting of superstructure reflections at room temperature for CSIO/STO. We noticed that the similar 4-subdomain monoclinic structure was reported recently for the epitaxial SrIrO$_3$ film and SrRuO$_3$ film grown on the STO substrate.[9,32] However, those studies were limited to room temperature only and no temperature-dependent measurements as shown below were carried out to investigate the structural evolution.

With increasing temperature, the splitting of (0, 1/2, 3) reflection along the H direction diminishes gradually, as shown in Fig. 3(c). Above 500 K, no splitting is observed and the profile can be fitted perfectly with a single peak, suggesting the disappearance of the twinning within domain B. From Fig. 3(c), the merging from 4 subdomains (see Fig. 3(e) and (f)) into 2 domains (see Fig. 3(g) and (h)) is estimated to occur around Ts = 510(5) K.

Our observation of the merging of the split superstructure peaks suggests the disappearance of the monoclinic distortions at 520 K, which means that a structural phase transition occurs at this temperature. As shown in Fig. 4(b), 2 orthorhombic domains (A and B) with all angles being 90° give rise to a single peak at either (1/2, 0, 3) or (0, 1/2, 3). Accordingly, the octahedral rotation pattern of the CSIO film layer has to be $a^+a^-c^0$ and $a^-a^+c^0$, with zero rotation of the $IrO_6$ octahedron around $c_{pc}$. These rotation patterns correspond to the space group of *Cmcm* (orthorhombic).

After clarifying the specific octahedral rotation patterns in the CSIO/STO film at both room temperature and 520 K, we can now conclude that the merging from 4 subdomains (A1, A2, B1, and B2) into 2 domains (A and B) around 510(5) K actually corresponds to a change of $IrO_6$ octahedral rotation pattern from $a^+a^-c^-$ to $a^+a^-c^0$ and a structural phase transition from monoclinic (space group $P112_1/m$) to orthorhombic (space group *Cmcm*) accordingly. With increasing temperature, the magnitude of octahedral rotation around the $c_{pc}$ axis vanishes gradually, corresponding to a change in the rotation pattern from $a^+a^-c^-$ to $a^+a^-c^0$ and the monoclinic-to-orthorhombic structural transition at $T_S$ = 510(5) K. Such a hidden structural phase transition is only visible in the thermal evolution of superstructure reflections, as it involves a subtle change in the octahedral rotation pattern but not that of the average lattice symmetry. At even higher temperatures, further structural transitions into higher-symmetry such as tetragonal and cubic phases are expected when the octahedral rotations around $a_{pc}$ and $b_{pc}$ axes disappear, as the case in many $ABO_3$ perovskites in bulk and thin-film forms.[33-35] Unfortunately, it is beyond the accessible temperature range of our experimental set-up.

In contrast, for the CSIO/GSO film, the K-scan profiles of superstructure reflection (1/2, 0, 3) can also be fitted with two components (as the dashed lines in Fig. 5(a) show), suggesting two twinned monoclinic subdomains (A1' and A2') with the octahedral rotation pattern of $a^+b^-c^-$ mirrored by the pseudocubic $ac$ plane (Fig. 5(b)). However, as (0, 1/2, 3) reflection is absent (see Fig. 2(d)), it is indicated that the $a^-b^+c^-$ domains mirrored by the $bc$ plane do not exist in the CSIO/GSO film layer. This 2-subdomain monoclinic structure is not surprising, considering the lower symmetry of the GSO substrate (orthorhombic [11]) compared with that of STO (cubic). As $a_{pc}$ and $b_{pc}$ of the CSIO film have to align with those of the GSO substrate, in which $a \neq b$, the monoclinic domains are deposited with some preferred orientation in the CSIO layer and only 2 domains are formed. Furthermore, the peak profiles of the (1/2, 0, 3) reflection do not show any visible difference between the room temperature and 520 K, indicating that a similar structural phase transition observed for CSIO/STO is absent. Therefore, the monoclinic phase with the octahedral rotation pattern of $a^+b^-c^-$ (space group $P112_1/m$) seems stable in the CSIO/GSO thin film for a very wide range of temperatures. Confined by the orthorhombic symmetry of the GSO substrate, which does not show any transformation into the tetragonal phase up to 1200℃,[36] the CSIO film grown on it does not display any structural instability with increasing temperature up to 520 K.

It is quite interesting that the CSIO film consists of multiple monoclinic domains when grown on top of the cubic STO substrate. Although it is inferred from the RSM shown in Fig. 2(a) that the global symmetry of the lattice is tetragonal, as a result of the compressive epitaxial strain and substrate clamping, the $IrO_6$ octahedron in the CSIO layer display a rotation pattern corresponding to a monoclinic and a orthorhombic space group, at room temperature and 520 K, respectively. Since the space group refers only to symmetry operations and not the shape of the unit cell itself, the existence of a unique phase with a monoclinic or an orthorhombic space group but a tetragonal lattice is fundamentally allowed for a film layer deposited on certain substrate, although it is unimaginable for a free-standing crystal.

By comparing the CSIO/STO and CSIO/GSO films, it is evident that the symmetry of the substrate is playing the key role in determining the structural phase of the film layer. Although previous studies have demonstrated that the rotation pattern of the $BO_6$ octahedra can change with the strength of epitaxial strain in the $ABO_3$ perovskite thin film, resulting in different possible structural phases,[37] the main driving force behind the structural transition in CSIO/STO is most likely the higher symmetry of the STO substrate instead of the epitaxial strain, as the strain in CSIO/STO is estimated to be even smaller than that in CSIO/GSO. Choosing suitable substrates with a specific symmetry will be undoubtedly of great importance for future synthesis of epitaxial thin films with desired structural properties.

## 4. Conclusion

In summary, we have investigated the structures of epitaxial $Ca_{0.5}Sr_{0.5}IrO_3$ films grown on $SrTiO_3$ (001) and $GdScO_3$ (110) substrates using high-resolution x-ray diffraction measurements. The CSIO/STO and CSIO/GSO films were found to show distinct structural properties, in aspects of global lattice symmetries, octahedral rotation patterns, domain formations and the tendencies towards structural instability. Although the CSIO/STO film displays a tetragonal lattice symmetry evidenced by the reciprocal space mapping, investigations of the superstructure reflections from the film layer suggests the space group of $P112_1/m$ (monoclinic) at room temperature due to $a^+a^-c^-$-type rotation of the $IrO_6$ octahedra. The monoclinic distortion of CSIO/STO film diminishes with increasing temperature, and a structural phase transition to the *Cmcm* space group (orthorhombic) with $a^+a^-c^0$ octahedral rotation occurs at $T_S = 510(5)$ K. In contrast, the CSIO/GSO film displays a stable monoclinic symmetry with $a^+b^-c^-$ octahedral rotation, showing no structural instability up to 520 K. Our study illustrates the importance of the symmetry of the substrate in the structure of epitaxial thin films and shed light on future engineering of heterostructures with desired structural properties.

**Acknowledgements**

Preliminary synchrotron x-ray diffraction measurements were performed at the X22 beamline at the National Synchrotron Light Source (NSLS). Work at the University of Toronto was supported by the Natural Sciences and Engineering Research Council of Canada, through Discovery and CREATE program, and Canada Foundation for Innovation. Use of the Canadian Light Source is supported by the Canada Foundation for Innovation, the Natural Sciences and Engineering Research Council of Canada, the University of Saskatchewan, the Government of Saskatchewan, Western Economic Diversification Canada, the National Research Council of Canada, and the Canadian Institutes of Health Research. This work was supported by the Institute for Basic Science (IBS) in Korea (IBS-R009-D1).

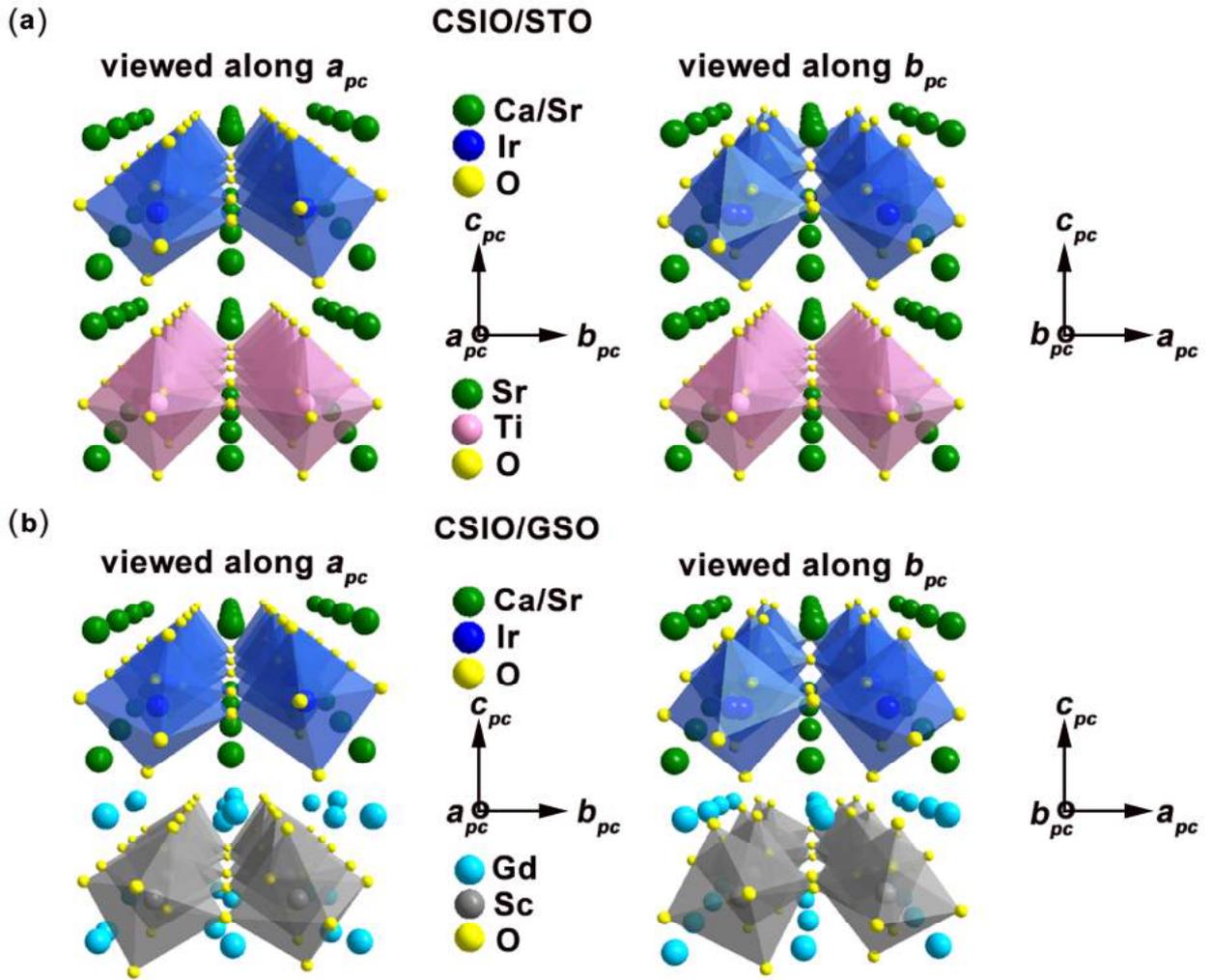

**Figure 1**. Orientations of the pseudocubic unit cell of CSIO grown on the STO (a) and GSO (b) substrates, respectively, with the octahedral rotation pattern of $a^+a^-c^-$ (a) and $a^+b^-c^-$ (b) viewed along the $a_{pc}$ and $b_{pc}$ axes. The octahedral rotation patterns of the STO and GSO substrates are $a^0a^0a^0$ and $a^+b^-b^-$, respectively, following the Glazer notation. For CSIO/GSO, $a_{pc}$, $b_{pc}$ and $c_{pc}$ align with the <0 0 1>, <1 -1 0> and <1 1 0> direction of the orthorhombic GSO substrate, respectively.

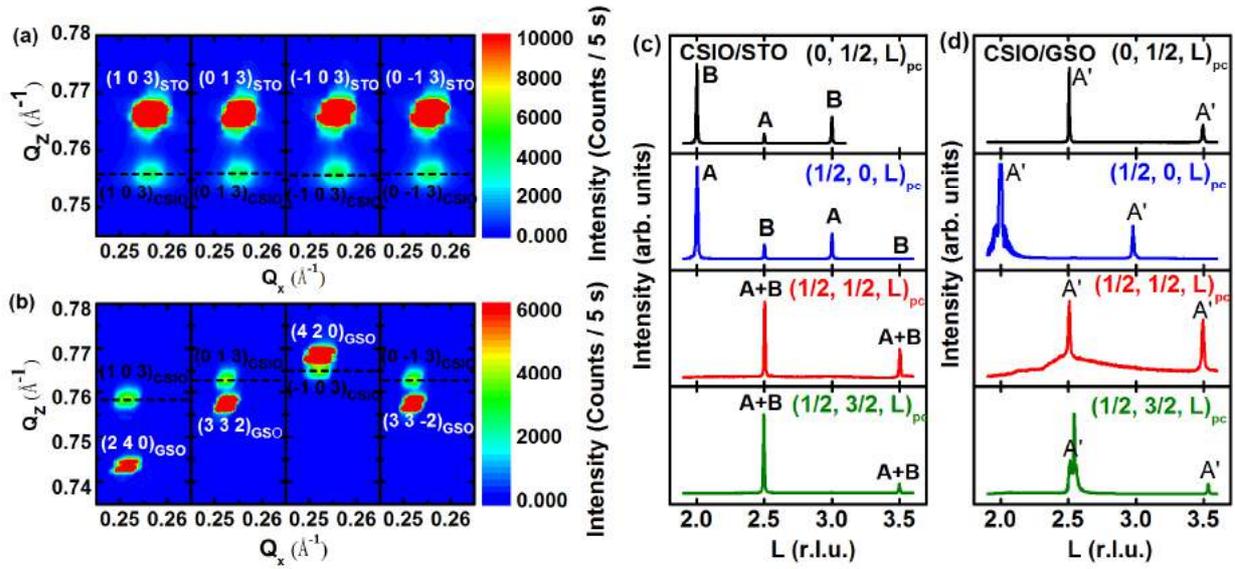

**Figure 2.** The reciprocal space maps of the CSIO/STO(a) and CSIO/GSO (b) films taken at room temperature, and the L-scans at room temperature revealing the half-integer superstructure reflections of the CSIO film layer grown on the STO (c) and GSO (d) substrates, respectively. The peaks marked by A and B in (c) arise from the $a^+a^-c^-$ and $a^-a^+c^-$ domains in CSIO/STO, respectively, while all peaks marked by A' in (d) are from a single $a^+b^-c^-$ domain in CSIO/GSO.

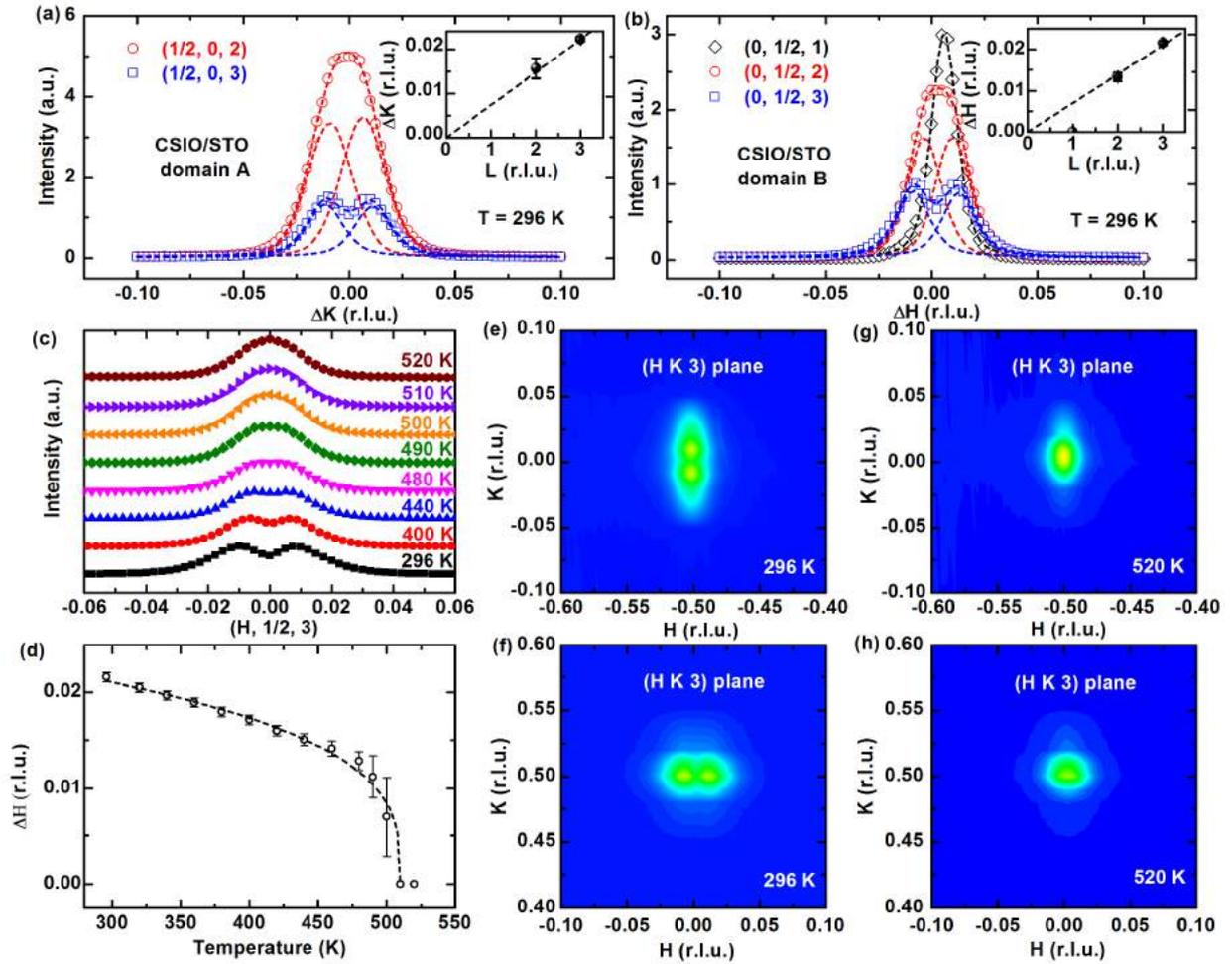

**Figure 3**. (a, b) Room-temperature K-scans of the CSIO/STO film across the superstructure peaks from domain A (a) and H-scans across those from domain B (b), respectively. The dashed lines represent the fitting to the profiles using the pseudo Voigt function. The L-dependencies of the peak splitting were plotted in the insets. (c) H-scans across the superstructure reflection (0, 1/2, 3) at different temperature. (d) The peak splitting as a function of temperature. The dashed line is a guide to the eyes. (e-h) Mesh scans around the (0, 1/2, 3) and (1/2, 0, 3) superstructure reflections at 296 K (e, f) and 520 K (g, h).

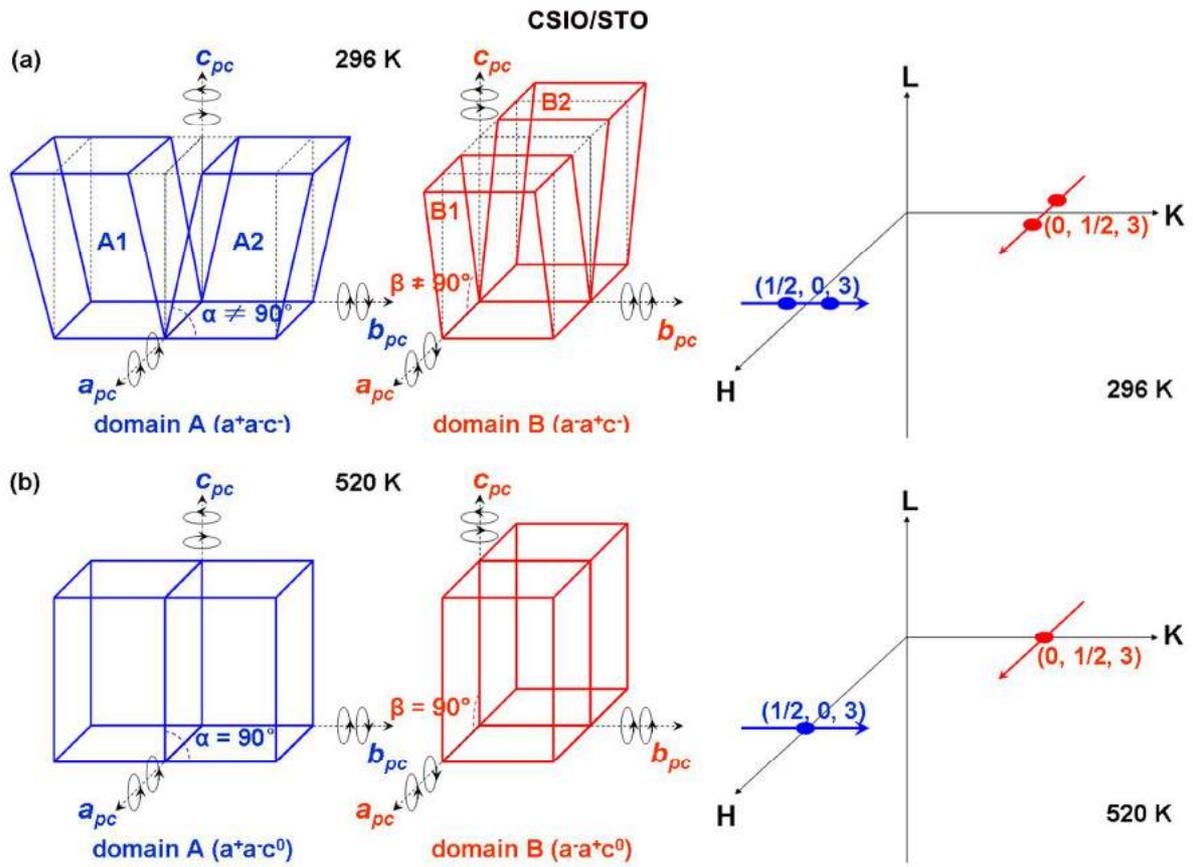

**Figure 4**. Schematic presentation of the domain structures and the corresponding (H, K, 3) plane in the reciprocal space of the CSIO/STO film at room temperature (a) and 520 K (b), respectively.

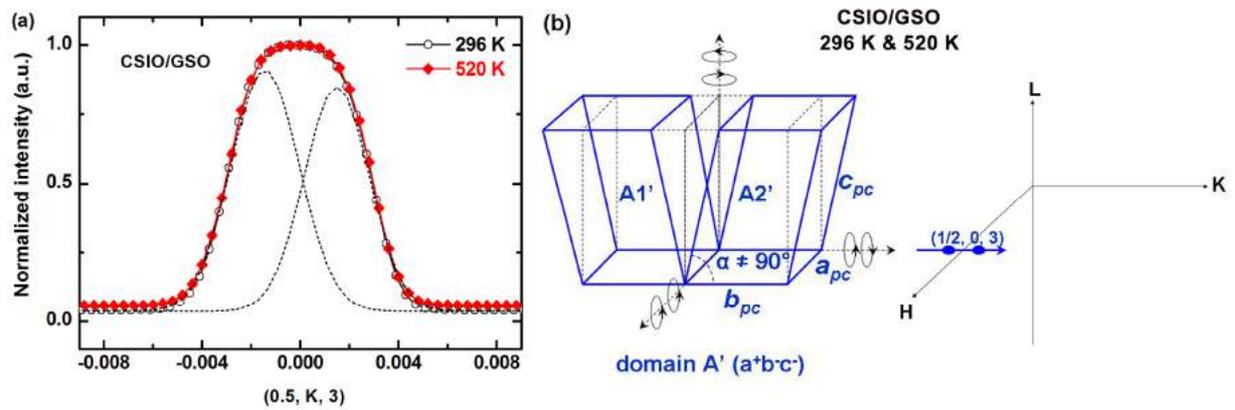

**Figure 5.** (a) K-scans of the superstructure peak (1/2, 0, 3) at room temperature and 520 K, respectively. The dashed lines show the fitting with two components to the profile at 296 K. (b) Schematic presentation of the domain structures and the corresponding (H, K, 3) plane in the reciprocal space of the CSIO/GSO film.